\documentclass[a4paper,11pt]{article}
\usepackage{pos}

\usepackage{physics,slashed,dsfont,amsmath,booktabs}

\newcommand{\Z}{\mathcal{Z}}
\newcommand{\D}{\mathcal{D}}
\newcommand{\ZT}{\mathrm{Z}(3)}
\newcommand{\SUT}{\mathrm{SU}(3)}

\title{First-order phase transition in dynamical 3-flavor QCD at imaginary isospin}

\author*[a,b]{Gergely Endr\H{o}di}
\author[c]{Guy D.\ Moore}
\author[d]{Alessandro Sciarra}

\affiliation[a]{Institute for Theoretical Physics,
ELTE E\"otv\"os Lor\'and University,\\
P\'azm\'any P.\ s\'et\'any 1/A, H-1117 Budapest, Hungary}

\affiliation[b]{Fakult\"at f\"ur Physik, Universit\"at Bielefeld,\\ Universit\"{a}tsstra{\ss}e 25, D-33615 Bielefeld, Germany}

\affiliation[c]{Institut f\"ur Kernphysik, Technische Universit\"at Darmstadt,\\ Schlossgartenstra{\ss}e 2,
D-64289 Darmstadt, Germany}

\affiliation[d]{Institut f\"{u}r Theoretische Physik, Goethe-Universit\"{a}t Frankfurt,\\
 Max-von-Laue-Str.\ 1, 60438 Frankfurt am Main, Germany}

\emailAdd{gergely.endrodi@ttk.elte.hu}
\emailAdd{guy.moore@physik.tu-darmstadt.de}
\emailAdd{sciarra@itp.uni-frankfurt.de}

\abstract{We revisit QCD with three mass-degenerate quark flavors at an imaginary isospin chemical potential set to $4\pi T/3$. This choice corresponds to a special point in the parameter space, where the theory possesses an exact $\mathrm{Z}(3)$ center symmetry. Through a finite-size scaling analysis, we demonstrate that in this case the finite temperature QCD transition is of first order and entails singular behavior both in the Polyakov loop and in the quark condensate. Our results are based on simulations with stout-smeared staggered quarks and a dedicated multi-histogram analysis.}

\FullConference{The 41st International Symposium on Lattice Field Theory (LATTICE2024)\\
 28 July - 3 August 2024\\
Liverpool, UK\\}

\begin{document}
\maketitle

\section{Introduction}

One of the most successful approaches to study QCD at nonzero density involves simulations at purely imaginary values of the chemical potentials, a region free of the complex action problem. QCD in the presence of imaginary chemical potentials also exhibits a rich phase structure that has non-trivial implications for the physics at real densities (see e.g.~\cite{DElia:2002tig,deForcrand:2010he,Bonati:2016pwz,Philipsen:2016hkv}). Some of these aspects may already be understood using perturbation theory -- and are exemplified by the well-known Roberge-Weiss phase transitions at high temperature and imaginary baryon density~\cite{Roberge:1986mm}.

The phase structure of QCD with three quark flavors can be described completely by three independent chemical potentials, i.e.\ an isospin $\mu_I$ and a strangeness $\mu_S$ chemical potential besides the baryon one $\mu_B$. The isospin axis is special in the sense that it is free of the complex action problem even for $\mu_I\in\mathds{R}$ and has enabled the precise determination of the phase diagram in the last years~\cite{Brandt:2017oyy,Brandt:2018omg}. The determination of the equation of state in this plane~\cite{Brandt:2022hwy,Abbott:2023coj,Abbott:2024vhj} is also of interest as it may for example serve as a non-trivial bound for the baryonic equation of state~\cite{Moore:2023glb,Fujimoto:2023unl}.

In this proceedings article, we consider three-flavor QCD at nonzero imaginary chemical potentials. This theory constitutes a generalization of the standard Roberge-Weiss setup and allows to explore the phase structure similarly to the two-flavor case~\cite{Brandt:2022jwo}. More importantly, for the specific choice $\mu_I=i4\pi T/3$, this system possesses an exact center symmetry~\cite{Kouno:2012zz,Cherman:2017tey} and is in general expected to exhibit a first-order deconfinement phase transition. Various aspects of this theory have already been investigated on the lattice~\cite{Iritani:2015ara,Misumi:2015hfa}, within effective QCD models~\cite{Kouno:2015sja,Li:2018xgl} and analytically using the 't Hooft anomaly~\cite{Tanizaki:2017qhf,Tanizaki:2017mtm}. Using a finite size scaling analysis and a multi-histogram reweighting method, we demonstrate that a first-order transition is indeed present and elaborate on its impact on fermionic observables and its dependence on the quark masses.

\section{Lattice setup and observables}

We consider the QCD partition function represented by the path integral for three flavors of rooted staggered quarks,
\begin{equation}
 \Z=\int \D U_\nu \, e^{-\beta S_g}\prod_{f=u,d,s} \det [\slashed{D}(\mu_f)+m]^{1/4}\,,
\end{equation}
where $S_g$ is the tree-level improved Symanzik action, $\beta=6/g^2$ the inverse gauge coupling and $\slashed{D}$ the twice stout-smeared staggered Dirac operator. Our lattices have geometry $N_s^3\times N_t$, corresponding to a physical volume $V=(N_sa)^3$ and a temperature $T=1/(N_ta)$, where $a$ is the lattice spacing. 
The masses of all three quarks are equal, $m\equiv m_u=m_d=m_s$ and run as a function of $\beta$ along the line of constant physics, which we take from full QCD at the physical point~\cite{Borsanyi:2010cj}. 
Most of our simulations are performed at the physical strange quark mass $m=m_s^{\rm phys}$, but we will also consider ligher and heavier quarks.

The chemical potentials for the three flavors are set to a purely imaginary isospin chemical potential $\mu_I=i 4\pi T/3$, corresponding to
\begin{equation}
\mu_u = i2\pi T/3, \qquad \mu_d=-i2\pi T/3, \qquad \mu_s=0\,.
\label{eq:imagiso_choice}
\end{equation}
For this particular choice, the 
theory is invariant under a simultaneous center transformation and a permutation of the three mass-degenerate quark flavors%
~\cite{Kouno:2012zz,Cherman:2017tey}. 
This fermionic theory therefore possesses an exact $\ZT$ center symmetry, just like pure gauge theory, for which the order parameter is the Polyakov loop,
\begin{equation}
 P=\frac{1}{V}\sum_\mathbf{n} \Tr \prod_{n_4=0}^{N_t-1} \widetilde U_4(\mathbf{n},n_4)\,,
\end{equation}
where $\widetilde U_\nu$ denote the twice stout-smeared gluon links that enter the Dirac operator.

Pure $\SUT$ gauge theory exhibits a first-order deconfinement phase transition associated with the spontaneous breaking of this center symmetry, where the Polyakov loop is discontinuous.
Similarly, the three-flavor system at the imaginary chemical potentials~\eqref{eq:imagiso_choice} is expected to exhibit a first-order deconfinement phase transition. 
Owing to the exact center symmetry, the expectation value of $P$ vanishes at any temperature. To characterize the transition, we therefore consider the `rotated' Polyakov loop $\bar P$~\cite{Iwasaki:1992ik}. This is obtained by realizing for each gauge configuration the center transformation that rotates the Polyakov loop to the real sector, $-\pi/3<\arg P \le \pi/3$, and taking its real part. Below we will also consider higher central moments $\kappa_n \equiv \langle (\bar P - \langle \bar P \rangle)^n\rangle$ of the rotated Polyakov loop distribution. These define the susceptibility, the skewness and the kurtosis of $\bar P$,
\begin{equation}
 \chi_{\bar P} = V \kappa_2, \qquad
 s_{\bar P} = \frac{\kappa_3}{\kappa_2^{3/2}}, \qquad
 B_{\bar P} = \frac{\kappa_4}{\kappa_2^2}\,.
 \label{eq:Pbarmoments}
\end{equation}


\section{Results}

\begin{figure}[b]
 \centering
 \mbox{
 \includegraphics[width=7cm]{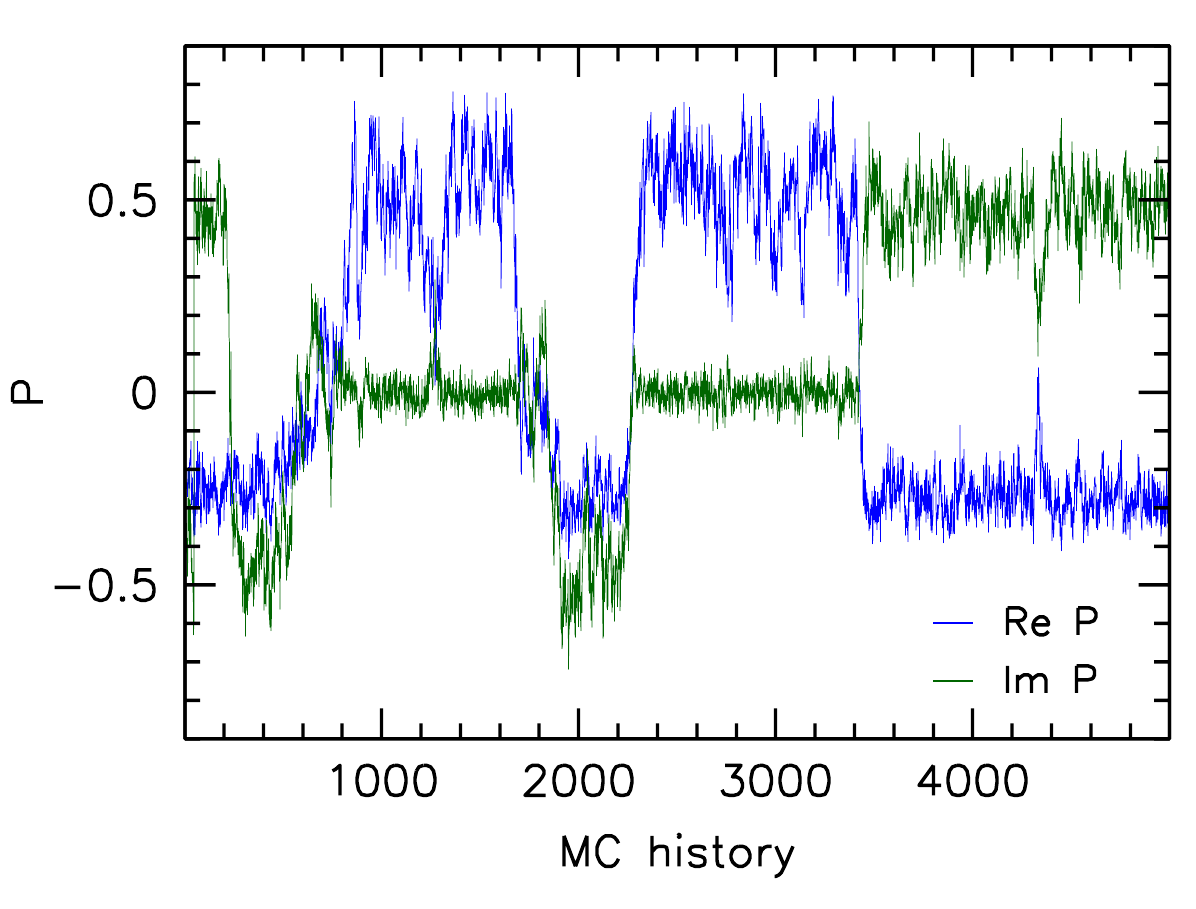}\quad
 \includegraphics[width=7cm]{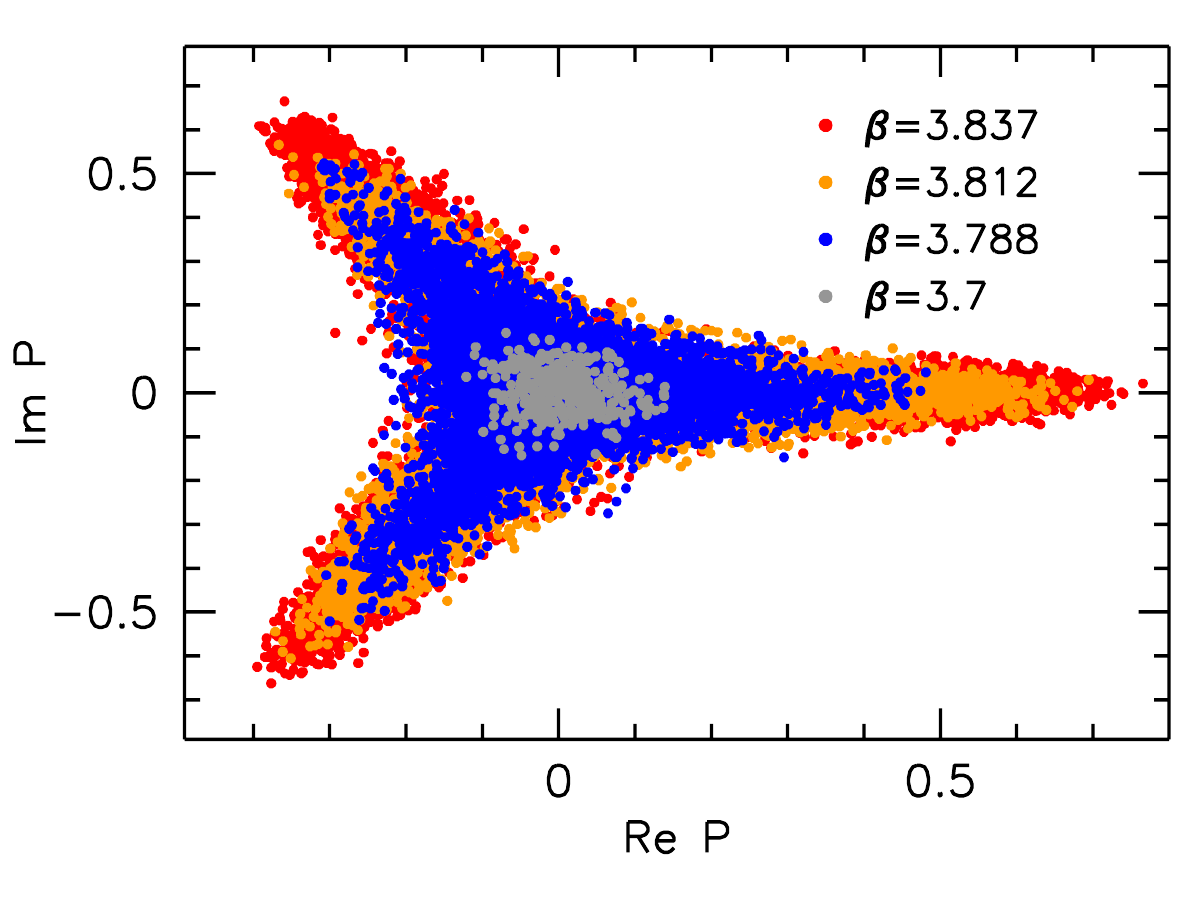}
 }
 \caption{
 Left panel: Monte Carlo history of the real (blue) and imaginary (green) part of the Polyakov loop on a $16^3\times 6$ ensemble at high temperature. Frequent jumps between the center sectors are observed. Right panel: scatter plot of the Polyakov loop in the complex plane on $16^3\times6$ ensembles at different temperatures. 
 \label{fig:ploop_hist}
 }
\end{figure}

We begin by presenting the Monte Carlo history of the (unrotated) Polyakov loop in the left panel of Fig.~\ref{fig:ploop_hist} on our $16^3\times6$ ensemble at $\beta=3.85$, corresponding to a high temperature. We can clearly observe frequent jumps between the three equivalent center sectors. This is further visualized in the right panel of the same figure, where the scatter plot of $P$ is shown in the complex plane. Here, different values of $\beta$, representing different temperatures, are compared. The figure demonstrates the abrupt change in the distribution, evolving from a center-symmetric vacuum around $P=0$ to a spontaneously broken distribution with the three equivalent center sectors. Notice that $\langle P\rangle=0$ holds for all temperatures.

\begin{figure}[t]
 \centering
 \mbox{
 \includegraphics[width=7cm]{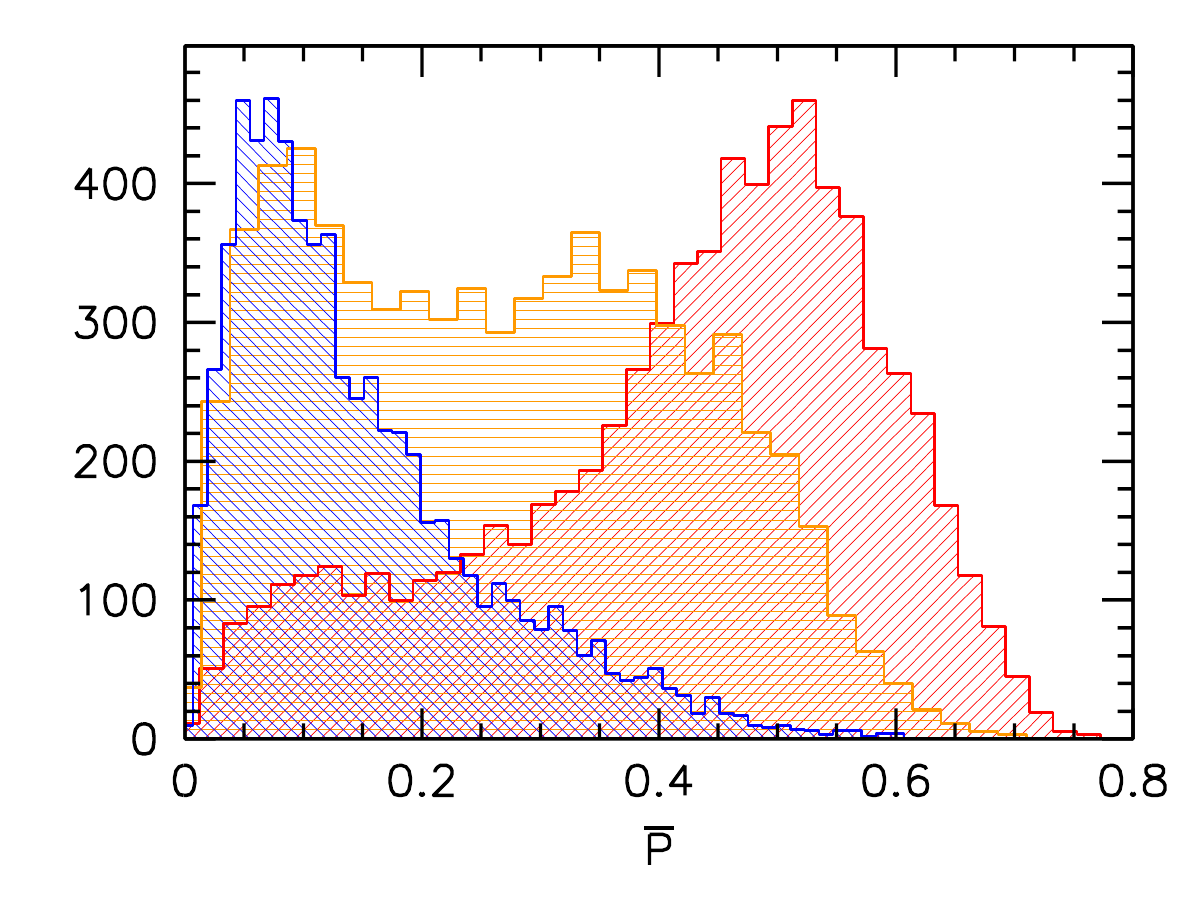}\quad
 \includegraphics[width=7cm]{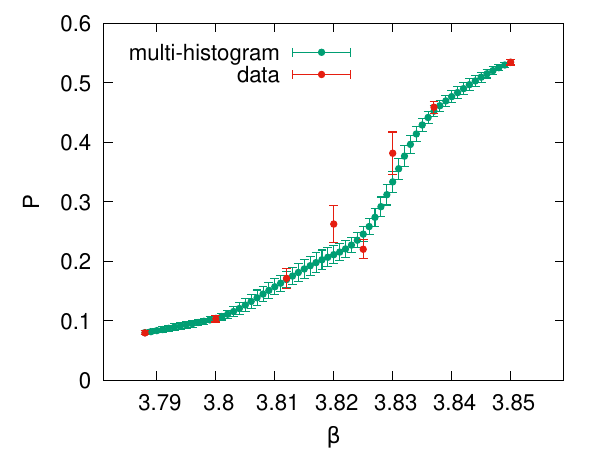}
 }
 \caption{
 Left panel: histogram of the rotated Polyakov loop on our $16^3\times6$ ensembles, with the same color coding for the temperatures as in the right panel of Fig.~\ref{fig:ploop_hist}.
 Right panel: temperature-dependence of the rotated Polyakov loop expectation value, together with an interpolation obtained using the multi-histogram method on our $20^3\times6$ lattices.
 \label{fig:Pbar_hist}
 }
\end{figure}

We proceed by showing the histogram of the rotated Polyakov loop in the left panel of Fig.~\ref{fig:Pbar_hist}, again for the $16^3\times6$ ensemble. The three colors represent $\beta=3.788$, $\beta=3.812$ and $\beta=3.837$, analogously to the colors used in the right panel of Fig.~\ref{fig:ploop_hist}. Here the transition is captured more transparently, with the histogram of $\bar P$ interpolating between the confined phase ($\bar P\approx 0$) and the deconfined one ($\bar P \neq 0$). Around the transition temperature, a double-peak structure starts to form, which becomes more prominent for larger volumes. At the same time, the mean $\bar P$ in the confined phase also approaches zero in the thermodynamic limit, which we will return to below.

The behavior of the histograms in the thermodynamic limit requires very large statistics. Here we focus instead on the moments~\eqref{eq:Pbarmoments}. In the right panel of Fig.~\ref{fig:Pbar_hist} we show the expectation value $\langle \bar P\rangle$ as a function of $\beta$. The identification of the transition is facilitated by an interpolation of our lattice data using the multi-histogram method~\cite{newmanb99}. This method corresponds to a weighted averaging of reweightings from each simulation point.  
Since in our simulations, the quark masses are tuned along the LCP, $m(\beta)$, the reweighting in $\beta$ also needs to be complemented by a reweighting in $m$. The latter we perform to leading order, i.e.\ we expand the logarithm of the reweighting factor to linear order in the quark mass difference. 
Explicitly, to reweight from a simulation point $\beta$ to a new point $\beta'$, the reweighting factor takes the form,
\begin{equation}
 W(\beta\to\beta') \approx \exp\left[ - (\beta'-\beta) \cdot \left( S_{g} - \frac{\partial m}{\partial \beta}\frac{V}{T}\, \bar\psi\psi\right) \right]\,,
\end{equation}
where $S_{g}$ and $\bar\psi\psi$ are the values of the gauge action and the quark condensate. The latter has the form
\begin{equation}
 \bar\psi\psi = \frac{1}{4} \frac{T}{V}\sum_{f=u,d,s} \Tr \left[\slashed{D}(\mu_f)+m\right]^{-1}\,,
 \label{eq:condensate}
\end{equation}
which is also invariant under center transformations~\cite{Cherman:2017tey}.
The function $\partial m/\partial \beta$ is taken from the LCP, determined at the physical point.

\begin{figure}[t]
 \centering
 \includegraphics[width=16cm,trim=0 0 0 20,clip]{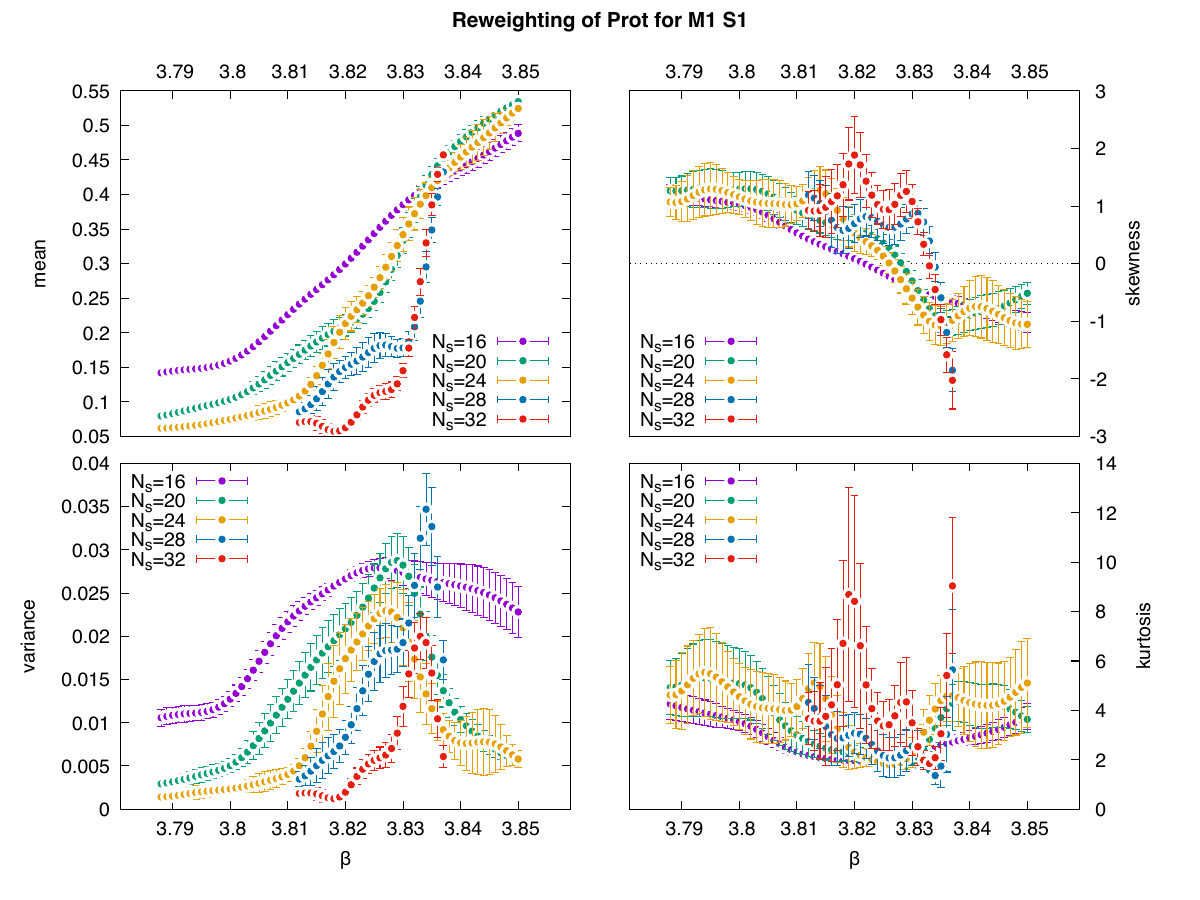}
 \caption{
 Moments of the distribution of the rotated Polyakov loop as a function of $\beta$ for different volumes. The panels show $\langle \bar P\rangle$ (top left), $\kappa_2$ (bottom left), $s_{\bar P}$ (top right) and $B_{\bar P}$ (bottom right).
 \label{fig:moments}
 }
\end{figure}

The multi-histogram method is particularly suited to reweighting higher moments of the $\bar P$ distribution in $\beta$ and therefore to more closely analyzing the transition. Fig.~\ref{fig:moments} shows the mean, the variance, the skewness and the kurtosis of the rotated Polyakov loop as a function of $\beta$ for different volumes ranging from $16^3\times 6 $ to $32^3\times 6$. 
The expectation value $\langle \bar P\rangle$ reveals a quick approach of the observable towards zero in the confined phase
and a characteristic sharpening of the transition as $V\to\infty$. The critical coupling is around $\beta_c\approx 3.835$, which we will determine below using the skewness.
The variance $\kappa_2$ also shows clear signatures of a real phase transition, featuring a narrower and narrower peak towards the thermodynamic limit. Note that the susceptibility is related to $\kappa_2$ as in Eq.~\eqref{eq:Pbarmoments}, therefore the fact that the peak height is roughly volume-independent translates to the scaling $\chi_{\bar P}\propto V$ for the susceptibility peak. This is a telltale sign of a first-order phase transition.

The skewness of the $\bar P$ distribution is also observed to feature an analogous sharpening and crosses zero at $\beta_c=3.83..(..)$, which we identify with the critical coupling. Finally, the kurtosis is found to have large uncertainties, so that we cannot make definite conclusions about its behavior. Nevertheless, $B_{\bar P}$ is observed to be consistent with unity at $\beta_c$. All in all, we observe strong indications for the first-order nature of the transition at this lattice spacing.

The behavior of fermionic observables around this deconfining phase transition is also of interest. Unlike pure gauge theory, the system at hand contains light fermions that also contribute to thermodynamics in a non-trivial way. Therefore, we investigate the quark condensate~\eqref{eq:condensate} next. Here, the reweighting becomes slightly more complicated. Note that since the $\bar\psi\psi$ operator itself depends on the quark mass, its leading-order reweighting from $\beta$ to $\beta'$ involves the evaluation of the operator at the shifted coupling. This is again performed in a leading-order expansion, which involves the connected susceptibility,
\begin{equation}
 \bar\psi\psi(\beta') \approx \bar\psi\psi(\beta) + (\beta'-\beta)\cdot\frac{\partial m}{\partial \beta} \chi_{\bar\psi\psi}, \qquad \chi_{\bar\psi\psi} = \frac{\partial \bar\psi\psi}{\partial m} = -\frac{1}{4}\frac{T}{V} \sum_{f=u,d,s} \Tr \left[\slashed{D}(\mu_f)+m\right]^{-2}\,.
\end{equation}

\begin{figure}[b]
 \centering
 \mbox{
 \includegraphics[width=7cm]{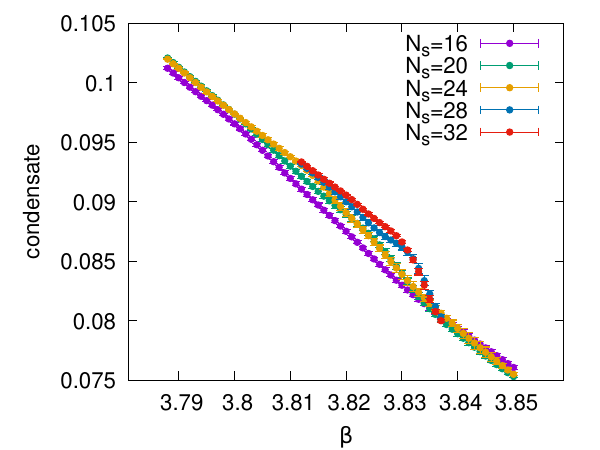}\quad
 \includegraphics[width=7cm]{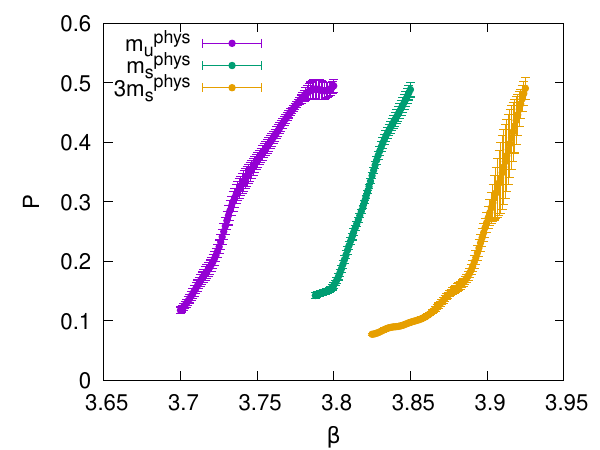}
 }
 \caption{
 Left panel: expectation value of the quark condensate as a function of $\beta$ for different volumes. Right panel: $\beta$-dependence of the rotated Polyakov loop on the $16^3\times6$ ensemble for three different quark masses.
 \label{fig:cond}
 }
\end{figure}

The left panel of Fig.~\ref{fig:cond} shows $\langle \bar\psi\psi\rangle$ interpolated in $\beta$ using the multi-histogram method for different volumes. Notice that this is a bare observable and is therefore subject to additive and multiplicative renormalization. Nevertheless, the imprint of the deconfinement phase transition is still visible in $\langle \bar\psi\psi\rangle$ via the emerging discontinuity at the critical gauge coupling as $V\to\infty$. 

The natural question to ask at this point is to what extent this behavior may be interpreted as a chiral transition. Our analysis so far was carried out at the physical strange quark mass $m=m_s^{\rm phys}$. The general expectation for the presence of a first-order phase transition is, however, independent of the actual quark mass value, as long as we have three degenerate flavors and the imaginary chemical potentials are assigned as in Eq.~\eqref{eq:imagiso_choice}. To assess the dependence of the results on the mass, we perform one further scan on the $16^3\times6$ ensemble at the physical light quark mass, $m=m_{ud}^{\rm phys}=m_s^{\rm phys}/28.15$~\cite{Borsanyi:2010cj} as well as at three times the physical strange quark mass, $m=3\cdot m_s^{\rm phys}$. The $\beta$-dependence of the rotated Polyakov loop is plotted in the right panel of Fig.~\ref{fig:cond}. Besides the qualitatively equivalent behavior in all cases, we observe a pronounced dependence $\beta_c(m)$.

In order to compare the critical temperatures in a transparent manner, we need to work in terms of dimensionless combinations. To this end, we choose the $w_0$ scale~\cite{BMW:2012hcm} determined from the Wilson flow~\cite{Luscher:2010iy}. We measured $w_0/a$ on $24^4$ lattices at zero chemical potentials and at the critical inverse gauge couplings. We find the values for $T_cw_0$ given in Tab.~\ref{tab:Tcw0}. It is instructive to compare these to the value of $T_cw_0=0.2507(2)$ in pure gauge theory~\cite{Borsanyi:2021yoz} as well as to the approximate value $T_cw_0\approx0.14$ that it takes at the crossover transition in physical $2+1$-flavor QCD~\cite{BMW:2012hcm}.

\begin{table}
	\centering
	\setlength{\tabcolsep}{10pt} 
	\begin{tabular}{*{4}{c}}
		\toprule
		 $m_{u,d,s}$ & $m_{ud}^{\rm phys}$ & $m_s^{\rm phys}$ & $3\cdot m_s^{\rm phys}$ \\
		 \midrule
		 $T_cw_0$ & 0.239(1) & 0.248(1) & 0.2505(5)  \\
		 \bottomrule
	\end{tabular}
	\caption{Measurements of the $w_0$ scale on $24^4$ lattices at $\mu_f=0$ at the critical couplings inferred from finite temperature $16^3\times6$ ensembles. \label{tab:Tcw0}}
\end{table}

\begin{figure}[b]
 \centering
 \mbox{
 \includegraphics[width=11cm]{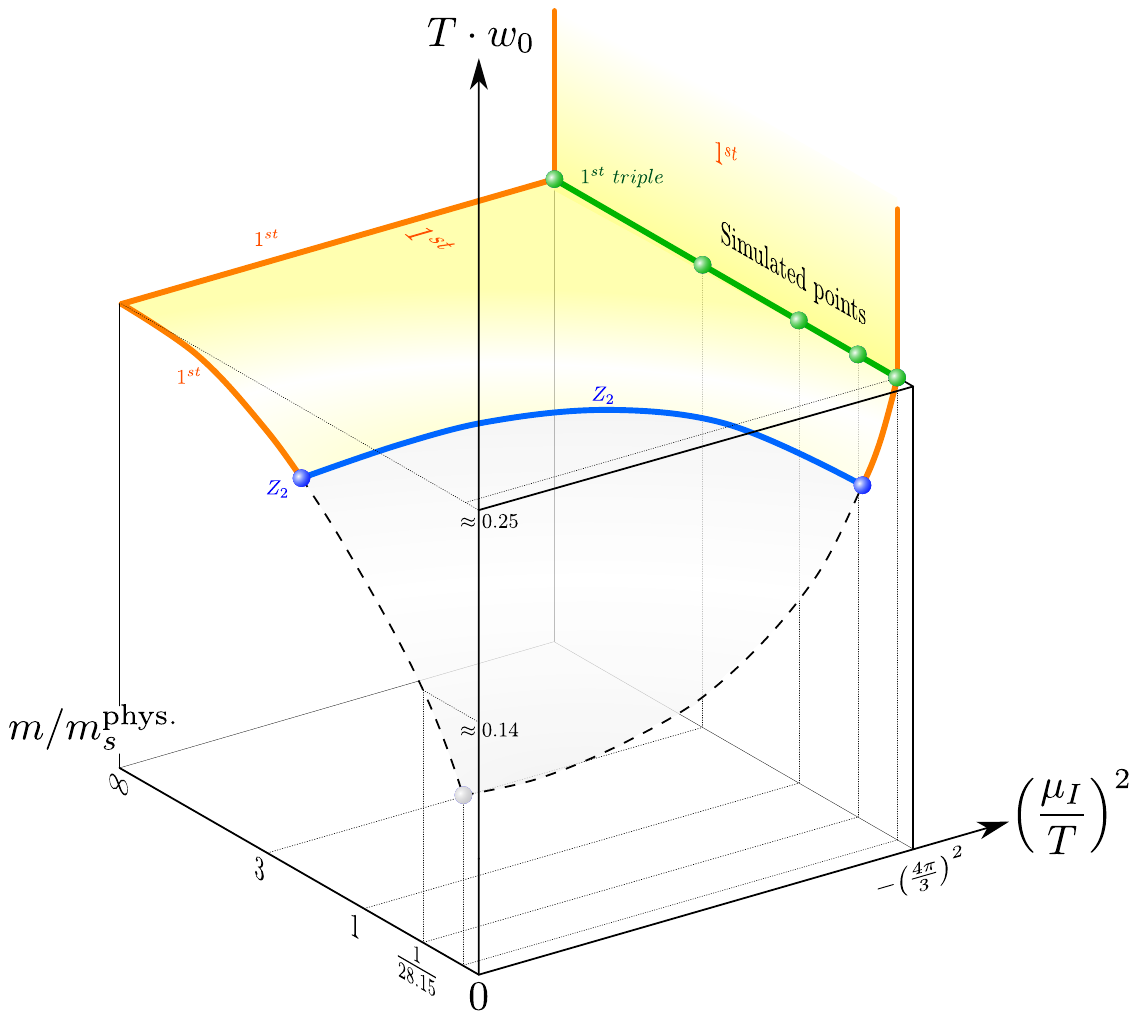}
 }
 \caption{
 Phase diagram of three-flavor QCD in the space spanned by the temperature normalized by $w_0$, the (squared) isospin chemical potential and the (degenerate) quark mass. The yellow area and orange lines indicate first-order phase transitions, the blue line a second-order $\mathrm{Z}(2)$ phase transition and the dashed lines crossovers. The vertical yellow area corresponds to a first-order transition separating deconfined phases with different Polyakov loop sectors, while the mostly horizontal area denotes the transition between confined and deconfined phases. Thus, the green line constitutes a line of first-order triple points, on which our simulation points also lie (green dots).
 \label{fig:phasediag}
 }
\end{figure}

We may collect all this information in the phase diagram of the theory. This is visualized in Fig.~\ref{fig:phasediag} in the temperature -- isospin chemical potential -- quark mass space. This three-dimensional phase diagram exhibits the exact center symmetry on both of its back-planes: the one at $\mu_I/T=i4\pi/3$ and the one at $m=\infty$.
The first-order deconfinement phase transition that we observed at $\mu_I/T=i4\pi /3$ and $m=m_s^{\rm phys}$ -- and expect to persist for practically any nonzero quark mass -- is therefore continuously connected to the first-order region at $\mu_I=0$ and high quark masses. The latter corresponds to the section of the diagonal in the upper right corner of the standard Columbia plot (see e.g.~\cite{Aarts:2023vsf}). This first-order transition terminates in a $\mathrm{Z}(2)$ second-order point, which becomes a $\mathrm{Z}(2)$ line in Fig.~\ref{fig:phasediag}. For even lower quark masses, the transition is an analytic crossover. Following Tab.~\ref{tab:Tcw0} and the above quoted values, we have also sketched the qualitative behavior of the transition temperature (in units of $1/w_0$) in the figure. 

Finally, we note that changing the imaginary isospin chemical potential in the deconfined, high-temperature phase at $\mu_I/T=i4\pi/3$ realizes yet another first-order phase transition. For $\mu_I$ smaller than this value, the preferred Polyakov loop sector is the real one (just like at $\mu_I=0$), while for slightly larger values the complex Polyakov loop sectors are favored and a remaining $\mathrm{Z}(2)$ symmetry is intact. The yellow vertical area in Fig.~\ref{fig:phasediag} therefore separates these two different deconfined phases. In the two-flavor theory, the associated Roberge-Weiss-type transitions have been mapped out recently~\cite{Brandt:2022iwk,Brandt:2022jwo}.

\section{Conclusions}

In this proceedings article we investigated three-flavor QCD with degenerate quark masses at a special value of the imaginary isospin chemical potential, where the theory exhibits an exact $\ZT$ center symmetry. We demonstrated that in this system, deconfinement is realized via a first-order phase transition. The analysis was facilitated by implementing a multi-histogram method for the interpolation of our observables, which also required a leading-order reweighting in the quark masses. The discontinuity in the Polyakov loop was shown to leave its imprint in the quark condensate and we argued that this behavior continues to hold for any nonzero quark mass values. Finally, we sketched the qualitative behavior of the QCD phase diagram in the mass-chemical potential-temperature space. Our results constitute a non-trivial example for a center-symmetric fermionic theory that exhibits a first-order phase transition.

\acknowledgments This work has received funding by the Deutsche Forschungsgemeinschaft (DFG, German Research Foundation) via CRC TRR 211 -- project number 315477589, the European Research Council (Consolidator Grant 101125637 CoStaMM) and the Hungarian National Research, Development and Innovation Office (Research Grant Hungary).

\bibliographystyle{JHEP.bst}
\bibliography{3fl_pos.bib}

\providecommand{\href}[2]{#2}\begingroup\raggedright\begin{thebibliography}{10}

\bibitem{DElia:2002tig}
M.~D'Elia and M.-P.~Lombardo, \emph{{Finite density QCD via imaginary chemical
  potential}}, \href{https://doi.org/10.1103/PhysRevD.67.014505}{\emph{Phys.
  Rev. D} {\bfseries 67} (2003) 014505}
  [\href{https://arxiv.org/abs/hep-lat/0209146}{{\ttfamily hep-lat/0209146}}].

\bibitem{deForcrand:2010he}
P.~de~Forcrand and O.~Philipsen, \emph{{Constraining the QCD phase diagram by
  tricritical lines at imaginary chemical potential}},
  \href{https://doi.org/10.1103/PhysRevLett.105.152001}{\emph{Phys. Rev. Lett.}
  {\bfseries 105} (2010) 152001}
  [\href{https://arxiv.org/abs/1004.3144}{{\ttfamily 1004.3144}}].

\bibitem{Bonati:2016pwz}
C.~Bonati, M.~D'Elia, M.~Mariti, M.~Mesiti, F.~Negro and F.~Sanfilippo,
  \emph{{Roberge-Weiss endpoint at the physical point of $N_f = 2+1$ QCD}},
  \href{https://doi.org/10.1103/PhysRevD.93.074504}{\emph{Phys. Rev. D}
  {\bfseries 93} (2016) 074504}
  [\href{https://arxiv.org/abs/1602.01426}{{\ttfamily 1602.01426}}].

\bibitem{Philipsen:2016hkv}
O.~Philipsen and C.~Pinke, \emph{{The $N_f=2$ QCD chiral phase transition with
  Wilson fermions at zero and imaginary chemical potential}},
  \href{https://doi.org/10.1103/PhysRevD.93.114507}{\emph{Phys. Rev. D}
  {\bfseries 93} (2016) 114507}
  [\href{https://arxiv.org/abs/1602.06129}{{\ttfamily 1602.06129}}].

\bibitem{Roberge:1986mm}
A.~Roberge and N.~Weiss, \emph{{Gauge Theories With Imaginary Chemical
  Potential and the Phases of {QCD}}},
  \href{https://doi.org/10.1016/0550-3213(86)90582-1}{\emph{Nucl. Phys. B}
  {\bfseries 275} (1986) 734}.

\bibitem{Brandt:2017oyy}
B.B.~Brandt, G.~Endr\H{o}di and S.~Schmalzbauer, \emph{{QCD phase diagram for
  nonzero isospin-asymmetry}},
  \href{https://doi.org/10.1103/PhysRevD.97.054514}{\emph{Phys. Rev. D}
  {\bfseries 97} (2018) 054514}
  [\href{https://arxiv.org/abs/1712.08190}{{\ttfamily 1712.08190}}].

\bibitem{Brandt:2018omg}
B.B.~Brandt and G.~Endr\H{o}di, \emph{{Reliability of Taylor expansions in
  QCD}}, \href{https://doi.org/10.1103/PhysRevD.99.014518}{\emph{Phys. Rev. D}
  {\bfseries 99} (2019) 014518}
  [\href{https://arxiv.org/abs/1810.11045}{{\ttfamily 1810.11045}}].

\bibitem{Brandt:2022hwy}
B.B.~Brandt, F.~Cuteri and G.~Endr\H{o}di, \emph{{Equation of state and speed
  of sound of isospin-asymmetric QCD on the lattice}},
  \href{https://doi.org/10.1007/JHEP07(2023)055}{\emph{JHEP} {\bfseries 07}
  (2023) 055} [\href{https://arxiv.org/abs/2212.14016}{{\ttfamily
  2212.14016}}].

\bibitem{Abbott:2023coj}
R.~Abbott, W.~Detmold, F.~Romero-L\'opez, Z.~Davoudi, M.~Illa, A.~Parre\~no
  et~al., \emph{{Lattice quantum chromodynamics at large isospin density: 6144
  pions in a box}},  \href{https://arxiv.org/abs/2307.15014}{{\ttfamily
  2307.15014}}.

\bibitem{Abbott:2024vhj}
R.~Abbott, W.~Detmold, M.~Illa, A.~Parre\~no, R.J.~Perry, F.~Romero-L\'opez
  et~al., \emph{{QCD constraints on isospin-dense matter and the nuclear
  equation of state}},  \href{https://arxiv.org/abs/2406.09273}{{\ttfamily
  2406.09273}}.

\bibitem{Moore:2023glb}
G.D.~Moore and T.~Gorda, \emph{{Bounding the QCD Equation of State with the
  Lattice}},  \href{https://arxiv.org/abs/2309.15149}{{\ttfamily 2309.15149}}.

\bibitem{Fujimoto:2023unl}
Y.~Fujimoto and S.~Reddy, \emph{{Bounds on the Equation of State from QCD
  Inequalities and Lattice QCD}},
  \href{https://arxiv.org/abs/2310.09427}{{\ttfamily 2310.09427}}.

\bibitem{Brandt:2022jwo}
B.B.~Brandt, A.~Chabane, V.~Chelnokov, F.~Cuteri, G.~Endr\H{o}di and
  C.~Winterowd, \emph{{Light Roberge-Weiss tricritical endpoint at imaginary
  isospin and baryon chemical potential}},
  \href{https://doi.org/10.1103/PhysRevD.109.034515}{\emph{Phys. Rev. D}
  {\bfseries 109} (2024) 034515}
  [\href{https://arxiv.org/abs/2207.10117}{{\ttfamily 2207.10117}}].

\bibitem{Kouno:2012zz}
H.~Kouno, Y.~Sakai, T.~Makiyama, K.~Tokunaga, T.~Sasaki and M.~Yahiro,
  \emph{{Quark-gluon thermodynamics with the Z(N(c)) symmetry}},
  \href{https://doi.org/10.1088/0954-3899/39/8/085010}{\emph{J. Phys. G}
  {\bfseries 39} (2012) 085010}.

\bibitem{Cherman:2017tey}
A.~Cherman, S.~Sen, M.~Unsal, M.L.~Wagman and L.G.~Yaffe, \emph{{Order
  parameters and color-flavor center symmetry in QCD}},
  \href{https://doi.org/10.1103/PhysRevLett.119.222001}{\emph{Phys. Rev. Lett.}
  {\bfseries 119} (2017) 222001}
  [\href{https://arxiv.org/abs/1706.05385}{{\ttfamily 1706.05385}}].

\bibitem{Iritani:2015ara}
T.~Iritani, E.~Itou and T.~Misumi, \emph{{Lattice study on QCD-like theory with
  exact center symmetry}},
  \href{https://doi.org/10.1007/JHEP11(2015)159}{\emph{JHEP} {\bfseries 11}
  (2015) 159} [\href{https://arxiv.org/abs/1508.07132}{{\ttfamily
  1508.07132}}].

\bibitem{Misumi:2015hfa}
T.~Misumi, T.~Iritani and E.~Itou, \emph{{Finite-temperature phase transition
  of $N_{f}=3$ QCD with exact center symmetry}},
  \href{https://doi.org/10.22323/1.251.0152}{\emph{PoS} {\bfseries LATTICE2015}
  (2016) 152} [\href{https://arxiv.org/abs/1510.07227}{{\ttfamily
  1510.07227}}].

\bibitem{Kouno:2015sja}
H.~Kouno, K.~Kashiwa, J.~Takahashi, T.~Misumi and M.~Yahiro,
  \emph{{Understanding QCD at high density from a Z$_3$-symmetric QCD-like
  theory}}, \href{https://doi.org/10.1103/PhysRevD.93.056009}{\emph{Phys. Rev.
  D} {\bfseries 93} (2016) 056009}
  [\href{https://arxiv.org/abs/1504.07585}{{\ttfamily 1504.07585}}].

\bibitem{Li:2018xgl}
X.-F.~Li and Z.~Zhang, \emph{{Roberge-Weiss transitions at different center
  symmetry breaking patterns in a $\mathbb{Z}_{3}$-QCD model}},
  \href{https://doi.org/10.1103/PhysRevD.100.074026}{\emph{Phys. Rev. D}
  {\bfseries 100} (2019) 074026}
  [\href{https://arxiv.org/abs/1812.01373}{{\ttfamily 1812.01373}}].

\bibitem{Tanizaki:2017qhf}
Y.~Tanizaki, T.~Misumi and N.~Sakai, \emph{{Circle compactification and
  \textquoteright{}t Hooft anomaly}},
  \href{https://doi.org/10.1007/JHEP12(2017)056}{\emph{JHEP} {\bfseries 12}
  (2017) 056} [\href{https://arxiv.org/abs/1710.08923}{{\ttfamily
  1710.08923}}].

\bibitem{Tanizaki:2017mtm}
Y.~Tanizaki, Y.~Kikuchi, T.~Misumi and N.~Sakai, \emph{{Anomaly matching for
  the phase diagram of massless $\mathbb{Z}_N$-QCD}},
  \href{https://doi.org/10.1103/PhysRevD.97.054012}{\emph{Phys. Rev. D}
  {\bfseries 97} (2018) 054012}
  [\href{https://arxiv.org/abs/1711.10487}{{\ttfamily 1711.10487}}].

\bibitem{Borsanyi:2010cj}
S.~Bors\'anyi, G.~Endr\H{o}di, Z.~Fodor, A.~Jakov\'ac, S.D.~Katz, S.~Krieg
  et~al., \emph{{The QCD equation of state with dynamical quarks}},
  \href{https://doi.org/10.1007/JHEP11(2010)077}{\emph{JHEP} {\bfseries 11}
  (2010) 077} [\href{https://arxiv.org/abs/1007.2580}{{\ttfamily 1007.2580}}].

\bibitem{Iwasaki:1992ik}
Y.~Iwasaki, K.~Kanaya, T.~Yoshie, T.~Hoshino, T.~Shirakawa, Y.~Oyanagi et~al.,
  \emph{{Finite temperature phase transition of SU(3) gauge theory on N(t) = 4
  and 6 lattices}}, \href{https://doi.org/10.1103/PhysRevD.46.4657}{\emph{Phys.
  Rev. D} {\bfseries 46} (1992) 4657}.

\bibitem{newmanb99}
M.E.J.~Newman and G.T.~Barkema, \emph{Monte Carlo methods in statistical
  physics}, Clarendon Press, Oxford (1999).

\bibitem{BMW:2012hcm}
{\scshape BMW} collaboration, \emph{{High-precision scale setting in lattice
  QCD}}, \href{https://doi.org/10.1007/JHEP09(2012)010}{\emph{JHEP} {\bfseries
  09} (2012) 010} [\href{https://arxiv.org/abs/1203.4469}{{\ttfamily
  1203.4469}}].

\bibitem{Luscher:2010iy}
M.~L\"uscher, \emph{{Properties and uses of the Wilson flow in lattice QCD}},
  \href{https://doi.org/10.1007/JHEP08(2010)071}{\emph{JHEP} {\bfseries 08}
  (2010) 071} [\href{https://arxiv.org/abs/1006.4518}{{\ttfamily 1006.4518}}].

\bibitem{Borsanyi:2021yoz}
S.~Bors\'anyi, Z.~Fodor, J.N.~Guenther, R.~Kara, P.~Parotto, A.~P\'asztor
  et~al., \emph{{The upper right corner of the Columbia plot with staggered
  fermions}}, \href{https://doi.org/10.22323/1.396.0496}{\emph{PoS} {\bfseries
  LATTICE2021} (2022) 496} [\href{https://arxiv.org/abs/2112.04192}{{\ttfamily
  2112.04192}}].

\bibitem{Aarts:2023vsf}
G.~Aarts et~al., \emph{{Phase Transitions in Particle Physics}: {Results and
  Perspectives from Lattice Quantum Chromo-Dynamics}},
  \href{https://doi.org/10.1016/j.ppnp.2023.104070}{\emph{Prog. Part. Nucl.
  Phys.} {\bfseries 133} (2023) 104070}
  [\href{https://arxiv.org/abs/2301.04382}{{\ttfamily 2301.04382}}].

\bibitem{Brandt:2022iwk}
B.B.~Brandt, V.~Chelnokov, F.~Cuteri, G.~Endr\H{o}di and C.~Winterowd,
  \emph{{The light Roberge-Weiss tricritical endpoint at imaginary isospin
  chemical potential}}, \href{https://doi.org/10.22323/1.396.0164}{\emph{PoS}
  {\bfseries LATTICE2021} (2022) 164}
  [\href{https://arxiv.org/abs/2207.10586}{{\ttfamily 2207.10586}}].

\end{thebibliography}\endgroup

\end{document}